\documentclass[twocolumn,showpacs,amsmath,amssymb,aps,prb,floatfix,nofootinbib,superscriptaddress]{revtex4}
\usepackage{graphicx,graphics,times}
\usepackage{bm}
\begin{document}
\title
{Etched Glass Surfaces, Atomic Force Microscopy and Stochastic
Analysis}
\author
{ G. R. Jafari $^{a,b}$, M. Reza Rahimi Tabar $^{c,d}$, A. Iraji zad
$^{c}$, G. Kavei $^f$}
\address
{\it $^a$ Department of Physics, Shahid Beheshti University, Evin,
Tehran 19839, Iran\\
$^b$ Department of Nano-Science, IPM,
P. O. Box 19395-5531, Tehran, Iran\\
$^c$  Department of Physics, Sharif University of
Technology, P. O. Box 11365-9161, Tehran, Iran \\
$^d$ CNRS UMR 6529, Observatoire de la C$\hat o$te d'Azur, BP
4229, 06304 Nice Cedex 4, France\\
 $^e$ Material and Energy,
Research Center, P.O. Box 14155-4777, Tehran, Iran}

\begin{abstract}

The effect of etching time scale of glass surface on its statistical
properties has been studied using atomic force microscopy technique.
We have characterized the complexity of the height fluctuation of a
etched surface by the stochastic parameters such as intermittency
exponents, roughness, roughness exponents, drift and diffusion
coefficients and find their variations in terms of the etching time.


\end{abstract}

\maketitle

\section{Introduction}

The complexity of rough surfaces is subject of a large variety of
investigations in different fields of science
\cite{Barabasi,Davies}. Surface roughness has an enormous influence
on many important physical phenomena such as contact mechanics,
sealing, adhesion, friction and self-cleaning paints and glass
windows, \cite{Bo,Zhao}. A surface roughness of just a few
nanometers is enough to remove the adhesion between clean and
(elastically) hard solid surfaces \cite{Bo}. The physical and
chemical properties of surfaces and interfaces are to a significant
degree determined by their topographic structure. The technology of
micro fabrication of glass is getting more and more important
because glass substrates are currently being used to fabricate micro
electro mechanical system (MEMS) devices \cite{Won}. Glass has many
advantages as a material for MEMS applications, such as good
mechanical and optical properties. It is a high electrical
insulator, and it can be easily bonded to silicon substrates at
temperatures lower than the temperature needed for fusion bonding
\cite{Melvin}. Also micro and nano-structuring of glass surfaces is
important for the production of many components and systems such as
gratings, diffractive optical elements, planar wave guide devices,
micro-fluidic channels and substrates for (bio) chemical
applications \cite{Cheng}. Wet etching is also well developed for
some of these applications
\cite{Knotter,Spierings,Schuitema,Glebov,Jafari1,Silikas,Irajizad}.

 One of the main problems in the rough surface is the scaling
behavior of the moments of height $h$ and  evolution of the
probability density function (PDF) of $h$, i.e. $P(h, x)$ in terms
of the length scale $x$. Recently some authors have been able to
obtain a Fokker-Planck equation describing the evolution of the
probability distribution function in terms of the length scale, by
analyzing some stochastic phenomena, such as rough surfaces
\cite{Jafari2,Waechter,Sangpour}, turbulent system \cite{Renner},
financial data \cite{Renner2},  cosmic background radiation
\cite{Ghasemi} and heart interbeats \cite{pei04} etc. They noticed
that the conditional probability density of field increment
satisfies the Chapman-Kolmogorov equation. Mathematically, this is
a necessary condition for the fluctuating data to be a Markovian
process in the length (time) scales \cite{Risken}. \vskip 1cm

 In this work,
we investigate the etching process as a stochastic process. We
measure the intermittency exponents of height structure function,
roughness, roughness exponents and Kramers-Moyal`s (KM)
coefficients. Indeed we consider the etching time $t$, as an
external parameter, to control the statistical properties of a
rough surface and find their variations with $t$. It is shown that
the first and second KM`s coefficients have well-defined values,
while the third and fourth order coefficients tend to zero. The
first and second KM`s coefficients for the fluctuations of $h(x)$,
enables us to explain the height fluctuation of the etched glass
surface.

\section{Experimental}

 We started with glass microscope slides as a sample. Only one
side of samples was etched by HF solution for different
 etching time (less than 20 minutes). HF concentration was $\%40$ for
all the experiments. The surface topography of the etched glass
samples in the scale ($< 5 \mu m$) was obtained using an AFM (Park
Scientific Instruments). The images in this scale were collected in
a constant force mode and digitized into $256$$\times$$256$ pixels.
A commercial standard pyramidal $Si_3N_4$ tip was used. A variety of
scans, each with size $L$, were recorded at random locations on the
surface. Figure 1 shows typical AFM image with resolutions of about
$20 nm$.

\section{Statistical quantities}

\subsection{Multifractal Analysis and the Intermittency Exponent }

 Assuming statistical translational invariance, the
structure functions $S^{q}(l)=<|h(x+l)-h(x)|^{q}>$, (moments of the
increment of the rough surface height fluctuation $h(x)$) will
depend only on the space deference of heights $l$, and has a power
law behavior if the process has the scaling property:
\begin{equation}
S^{q}(l)=<|h(x+l)-h(x)|^{q}> \propto
S^{q}(L_{0})(\frac{l}{L_{0}})^{\xi(q)}
\end{equation}
where $L_{0}$ is the fixed largest length scale of the system, $<
\cdot \cdot \cdot>$ denotes statistical average (for non-overlapping
increments of length $l$), $q$ is the order of the moment (we take
here $q > 0$), and $\xi(q)$ is the exponents of structure function.
The second moment is linked to the slope $\beta$ of the Fourier
power spectrum: $\beta=1+\xi_{2}$. The main property of a
multifractal processes is that it is characterized by a non-linear
$\xi_{q}$ function verses $q$. Monofractals are the generic result
of this linear behavior. For instance, for Brownian motion (Bm)
$\xi_{q} = q/2$, and for fractional Brownian motion (fBm) $\xi_{q}
\propto q$.
\begin{figure}
\includegraphics[width=8truecm]{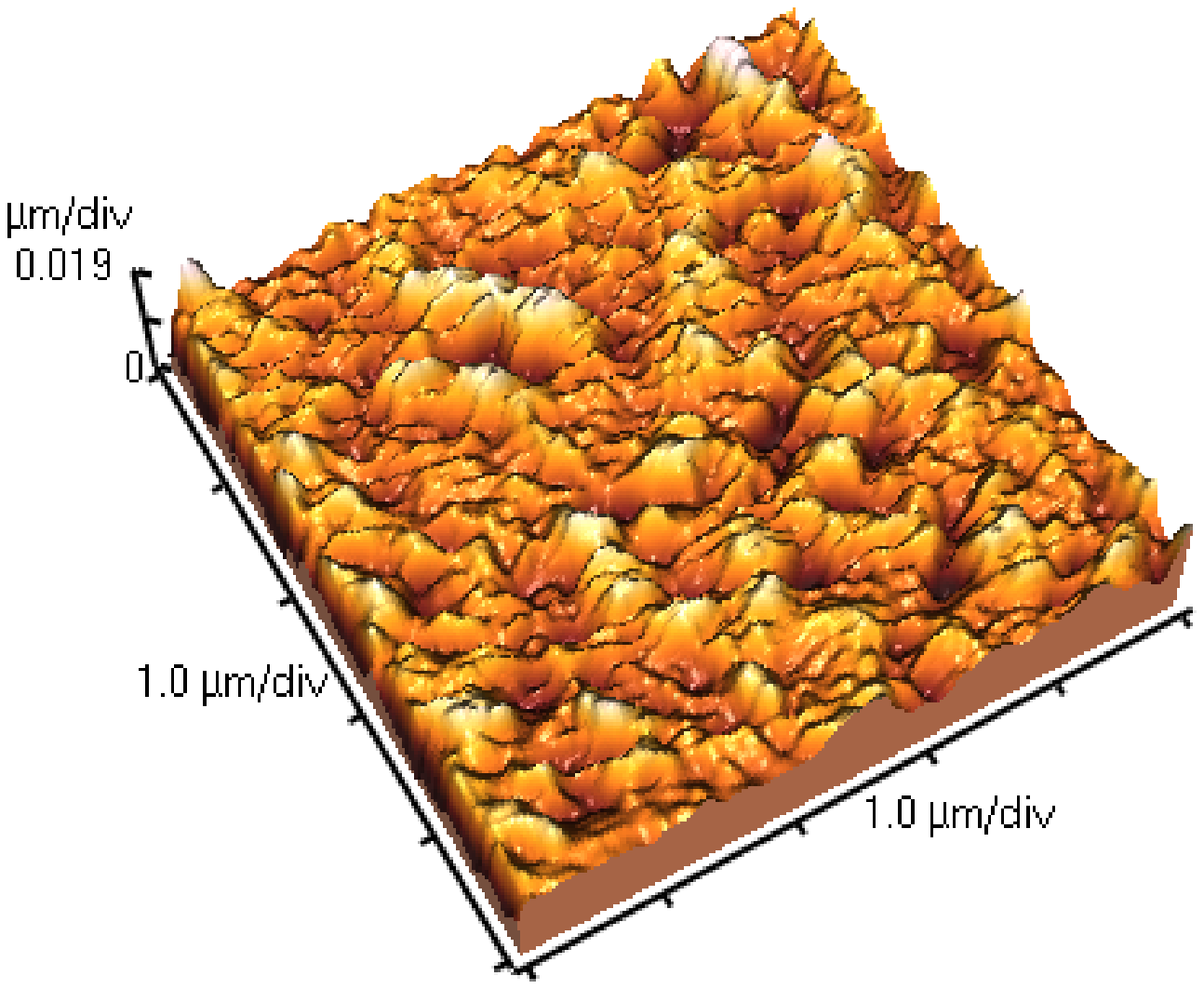}
 \narrowtext \caption{AFM surface image of etched glass
film with size  $5\times 5 \mu m^2 $  after 12 minutes.}
\end{figure}

\subsection {Roughness and Roughness Exponents}

It is also known that to derive the quantitative information of the
surface morphology one may consider a sample of size $L$ and define
the mean height of growing film $\overline{h}$ and its {\it
variance}, $\sigma$ by:
\begin{equation}\label{w}
\sigma(L,t) =(\langle (h-\overline{h})^2\rangle)^{1/2}
\end{equation}
where $t$ is etching time and $\langle\cdots\rangle$ denotes an
averaging over different samples, respectively. Moreover, etching
time is a factor which can apply to control the surface roughness of
thin films.

 Let us now calculate also the roughness exponent of the
etched glass. Starting from a flat interface (one of the possible
initial conditions), it is conjectured that a scaling of space by
factor $b$ and of time by factor $b^z$ ($z$ is the dynamical scaling
exponent), rescales the variance, $\sigma$ by factor $b^{\chi}$ as
follows \cite{Barabasi}:
\begin{equation}\label{scaling}
\sigma(bL,b^zt)=b^{\alpha}\sigma(L,t)
\end{equation}
which implies that
\begin{equation}
\sigma(L,t)=L^{\alpha}f(t/L^z).
\end{equation}
If for large $t$ and fixed $L$ $(x=t / L^z \rightarrow \infty)$
$\sigma$ saturate.  However, for fixed large $L$ and $t\ll L^z$, one
expects that correlations of the height fluctuations are set up only
within a distance $t^{1/z}$ and thus must be independent of $L$.
This implies that for $x \ll 1$, $f(x)\sim x^{\beta}$ with
$\beta=\alpha / z$. Thus dynamic scaling postulates that
\begin{eqnarray}
\sigma(L,t)\propto
\left\{%
\begin{array}{ll}
   t^{\beta}, & \hbox{t$\ll L^z$;}\\
   L^{\alpha}, & \hbox{t$\gg L^{z}$}. \\
\end{array}%
\right.
\end{eqnarray}
The roughness exponent $\alpha$ and the dynamic exponent $\beta$
characterize the self-affine geometry of the surface and its
dynamics, respectively.

The common procedure to measure the roughness exponent
 of a rough surface is use of the surface structure function depending
on the length scale $l$ which is defined as:
\begin{eqnarray}\label{Structure}
S^{2}(l)=\langle|h(x+l)-h(x)|^2\rangle.
\end{eqnarray}
It is equivalent to the statistics of height-height correlation
function $C(l)$ for stationary surfaces, i.e.
$S^{2}(l)=2\sigma^2(1-C(l))$. The second order structure function
$S(l)$, scales with $l$ as $ l^{2\alpha}$ \cite{Barabasi}.

\subsection{The Markov Nature of Height Fluctuations: Drift and
Diffusion Coefficients}

We check whether the data of height fluctuations follow a Markov
chain and, if so, measure the Markov length scale $l_M$. As is
well-known, a given process with a degree of randomness or
stochasticity may have a finite or an infinite Markov length scale
\cite{Markov}. The Markov length scale is the minimum length
interval over which the data can be considered as a Markov process.
To determine the Markov length scale $l_M$, we note that a complete
characterization of the statistical properties of random
fluctuations of a quantity $h$ in terms of a parameter $x$ requires
evaluation of the joint PDF, i.e. $P_N(h_1,x_1;....;h_N,x_N)$, for
any arbitrary $N$. If the process is a Markov process (a process
without memory), an important simplification arises. For this type
of process, $P_N$ can be generated by a product of the conditional
probabilities $P(h_{i+1},x_{i+1}|h_i,x_i)$, for $i=1,...,N-1$. As a
necessary condition for being a Markov process, the
Chapman-Kolmogorov equation,
\begin{eqnarray}
  &&P(h_2,x_2|h_1,x_1)= \cr \nonumber\\
  &&\int \hbox{d} (h_i)\,
  P(h_2,x_2|h_i,x_i)\, P(h_i,
  x_i| h_1,x_1)
\end{eqnarray}
should hold for any value of $x_i$, in the interval $ x_2<x_i<x_1$
\cite{Risken}.

The simplest way to determine $l_M$ for homogeneous surface is the
numerical calculation of the quantity,
$S=|P(h_2,x_2|h_1,x_1)-\int\hbox{d}
h_3P(h_2,x_2|h_3,x_3)\,P(h_3,x_3|h_1,x_1)|$, for given $h_1$ and
$h_2$, in terms of, for example, $x_3-x_1$ and considering the
possible errors in estimating $S$. Then, $l_M=x_3-x_1$ for that
value of $x_3-x_1$ such that, $S=0$ \cite{Markov}.

 It is well-known, the Chapman-Kolmogorov
equation yields an evolution equation for the change of the
distribution function $P(h,x)$ across the scales $x$. The
Chapman-Kolmogorov equation formulated in differential form yields a
master equation, which can take the form of a Fokker-Planck equation
\cite{Risken,Markov}:
\begin{eqnarray}\label{Fokker}
  \frac {\partial}{\partial x} P(h,x)=
 [-\frac{\partial }{\partial h}
  D^{(1)}(h,x)
  +\frac{\partial^2 }{\partial h^2} D^{(2)}(h,x)]
  P(h,x).
\end{eqnarray}
The drift and diffusion coefficients $D^{(1)}(h, r)$, $D^{(2)}(h,
r)$ can be estimated directly from the data and the moments
$M^{(k)}$ of the conditional probability distributions:
\begin{eqnarray}\label{D(k)}
  && D^{(k)}(h,x) = \frac{1}{k!}
   {\rm lim}_{r \rightarrow 0}  M^{(k)} \cr \nonumber \\
  && M^{(k)} = \frac{1}{r}  \int dh'
  (h'-h)^k P(h',x+r|
  h,x).
\end{eqnarray}
\begin{figure}
\includegraphics[width=8truecm]{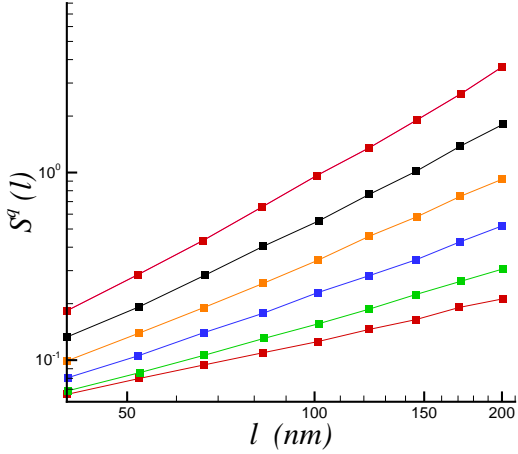}
 \narrowtext \caption{
Scaling of the structure functions in log-log plot for moments less
than 8. (from bottom to top).}
\end{figure}
\begin{figure}
\includegraphics[width=8truecm]{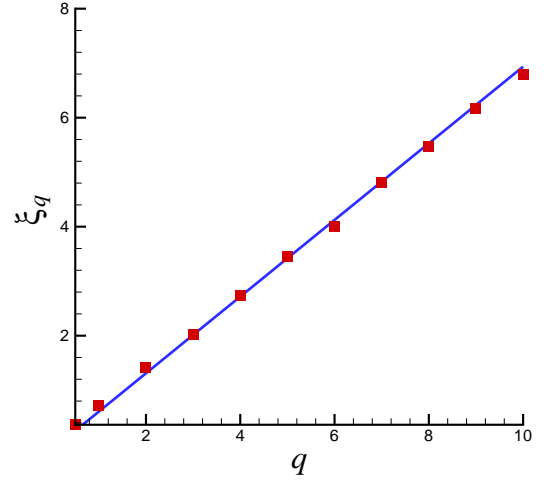}
 \narrowtext \caption{
The results of scaling exponent $\xi_{q}$ which is clearly linear
vs. q.}
\end{figure}
The coefficients $D^{(k)}(h,x)$`s are known as Kramers-Moyal
coefficients.  According to Pawula`s theorem \cite{Risken}, the
Kramers-Moyal expansion stops after the second term, provided that
the fourth order coefficient $D^{(4)} (h,x)$ vanishes \cite{Risken}.
 The forth order coefficients $D^{(4)}$
 in our analysis was found to be about $ {D^{(4)}} \simeq 10^{-4} {D^
{(2)}}$. In this approximation, we can ignore the coefficients
$D^{(n)}$ for $n \geq 3$. We note that this Fokker-Planck equation
is equivalent to the following Langevin equation (using the Ito
interpretation) \cite{Risken}:
\begin{equation}\label{Langevin}
  \frac{\partial}{\partial x}  h(x)=D^{(1)}(h,x) +
  \sqrt{D^{(2)}(h,x)}f(x)
\end{equation}
where $f(x)$ is a random force, zero mean with gaussian statistics,
$\delta$-correlated in $x$, i.e. $\langle
f(x)f(x')\rangle=2\delta(x-x')$. Furthermore, with this last
expression, it becomes clear that we are able to separate the
deterministic and the noisy components of the surface height
fluctuations in terms of the coefficients $D^{(1)}$ and $D^{(2)}$.
\begin{figure}
\includegraphics[width=8truecm]{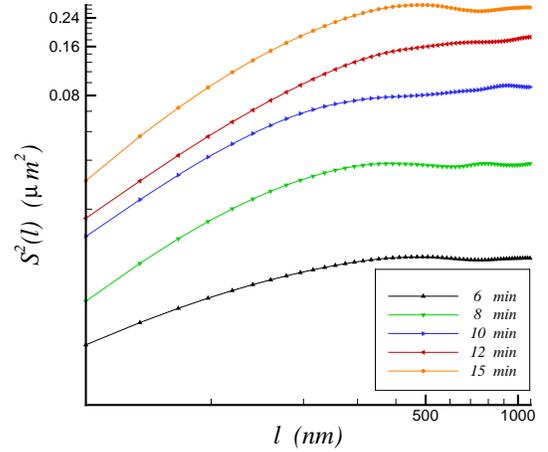}
\narrowtext \caption{ Log-Log
plot of selection structure function of the etched glass surfaces.}
\end{figure}

\section{Results and Discussion}

Now, using the introduced statistical parameters in the previous
sections, it is possible to obtain some quantitative information
about the effect of etching time on surface topography of the glass
surface. To study the effect of the etching time on the surface
statistical characteristics, we have utilized AFM  imaging technique
in order to obtain microstructural data of the etched glass surfaces
at the different etching time in the HF. Figure 1 shows the AFM
image of etched glass after $12$ minuets etched. To investigate the
scaling behavior of the moments of $\delta h_{l}=h(x+l)-h(x)$, we
consider the samples that they reached to the stationary state. This
means that their statistical properties do not change with time. In
our case the samples with etching time more than $20$ minutes are
almost stationary. Figure 2 shows the log-log plot of the structure
functions verses length scale $l$ for different orders of moments.
The straight lines show that the moments of order $q$ have the
scaling behavior. We have checked the scaling relation up to moment
$q=10$.  The resulting intermittency exponent $\xi_{q}$ is shown in
figure 3. It is evident that $\xi_{q}$ has a linear behavior. This
means that the height fluctuations are mono-fractal behavior. We
also directly estimated the scaling exponent of the linear term
$l^{qH}/<(h(x+l)-h(x))^{q}>$ and obtain the following values for the
samples with 20 minuets etching time, $\xi_{1} = 0.70 \pm 0.04$ and
$\xi_{2}= 1.40 \pm 0.04$. This means etching memorize fractal
feature during etching. Therefore using the scaling exponent
$\xi_{2}$ we obtain the roughness exponent $\alpha$ as $ \xi_{2}/2=
0.70 \pm 0.04$.
\begin{figure}
\begin{center}
\includegraphics[width=8truecm]{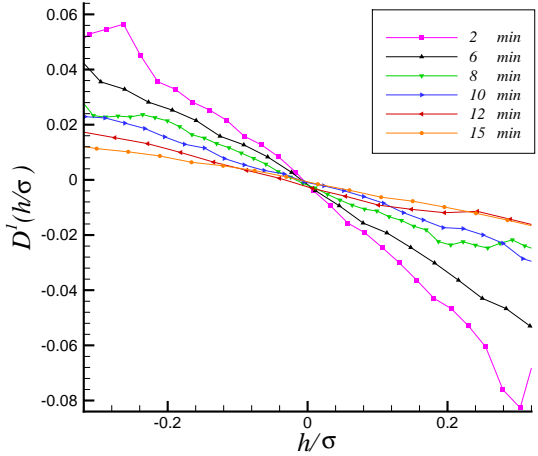}
 \narrowtext \caption{ Drift
coefficients of the surfaces at different etching time less than 20
minutes.}
\end{center}
\end{figure}
\begin{figure}
\begin{center}
\includegraphics[width=8truecm]{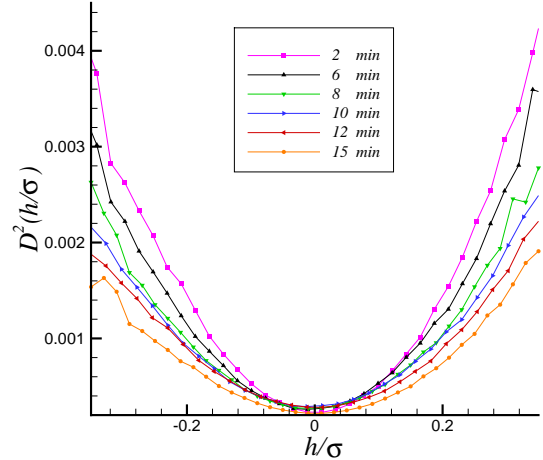}
\narrowtext \caption{ Diffused
coefficients of the surface at different etching time less than 20
minutes.}
\end{center}
 \end{figure}
 Figure 4 presents the structure function $S(l)$ of
the surface at the different etching time, using equation
(\ref{Structure}). It is also possible to evaluate the grain size
dependence to the etching time, using the correlation length
achieved by the structure function represented in figure 4. The
correlation lengths increase with etching time. Its value has a
exponential behavior $448(1-\exp(-0.15t))nm$. Also we find that the
dynamical exponent is given by $\beta = 0.6\pm 0.1$. Also we
measured the variation of the Markov length with etching time $t$
(min), and obtain $l_{M}=40+3t$ (nm) for time scales $t<20 ~min$.

Finally to obtain the stochastic equation of the height fluctuations
behavior of the surface, we need to measure the Keramer- Moyal
Coefficients. In our analysis the forth order coefficients $D^{(4)}$
is less than Second order coefficients, $D^{(2)}$, about $ {D^{(4)}}
\simeq 10^{-4} {D^ {(2)}}$. In this approximation, we ignore the
coefficients $D^{(n)}$ for $n \geq 3$. So, to discuss the surfaces
it just needs to measure the drift coefficient
$D^{(1)}(\frac{h}{\sigma})$ and diffusion coefficient
$D^{(2)}(\frac{h}{\sigma})$ using Eq. (\ref{D(k)}). Figures 5 and 6
show the drift coefficient $D^{(1)}(\frac{h}{\sigma})$ and diffusion
coefficients $D^{(2)}(\frac{h}{\sigma})$ for the surfaces at the
different etching time, respectively. It can be shown that the drift
and diffusion coefficients have the following behavior,
\begin{eqnarray}\label{D1}
D^{(1)} (\frac{h}{\sigma},t)=-f^{(1)}(t)\frac{h}{\sigma}
\end{eqnarray}
\begin{eqnarray}\label{D2}
D^{(2)} (\frac{h}{\sigma},t)=f^{(2)}(t)(\frac{h}{\sigma})^2
\end{eqnarray}
The two coefficients $f^{(1)}(t)$ and $f^{(2)}(t)$ increase with the
$\frac{h}{\sigma}$ then is saturated. Using the data analysis we
obtain that they are linear verses time (min): $f^{(1)}(t)=0.005 t$
and $f^{(2)}(t)=0.0003 t$ for time scales $t<20 ~min$. To better
comparing the parameter of samples we divided the heights to their
variances.  In this case, maximum and minimum of heights are about
plus 1 and mines 1, respectively. Comparing samples with etching
times 2 and 6 minutes, shows $f^{(1)}$ increases 300 percent after 4
minutes (from 2 min to 6 min) from $f^{(1)}(t=2 \times 60)=0.6$ to
$f^{(1)}(t=6 \times 60)=1.8$. Also, $f^{(2)}$ is $0.006$ and $0.018$
after 2 and 6  minutes, respectively.

\section{Conclusions}

We have investigated the role of etching time, as an external
parameter, to control the statistical properties of a rough
surface.  We have shown that in the saturate state the structure
of topography has fractal feature with fractal dimension
$D_{f}=1.30$. In addition, Langevin characterization of the etched
surfaces enable us to regenerate the rough surfaces grown at the
different etching time, with the same statistical properties in
the considered scales \cite{Jafari2}.

\section{Acknowledgment}

We would like to thank S. M. Mahdavi for his useful comments and
discussions and Also P. Kaghazchi and M. Shirazi for samples
preparation.

\end{document}